# Shallow and Deep Networks Intrusion Detection System: A Taxonomy and Survey

Elike Hodo, Xavier Bellekens, Andrew Hamilton, Christos Tachtatzis and Robert Atkinson

Department of Electronic & Electrical Engineering

University of Strathclyde

Division of Computing and Mathematics

University of Abertay Dundee

E-mail: elike.hodo@strath.ac.uk

E-mail: x.Bellekens@abertay.ac.uk

*Abstract*— Intrusion detection has attracted a considerable interest from researchers and industries. The community, after many years of research, still faces the problem of building reliable and efficient IDS that are capable of handling large quantities of data, with changing patterns in real time situations. The work presented in this manuscript classifies intrusion detection systems (IDS). Moreover, a taxonomy and survey of shallow and deep networks intrusion detection systems is presented based on previous and current works. This taxonomy and survey reviews machine learning techniques and their performance in detecting anomalies. Feature selection which influences the effectiveness of machine learning (ML) IDS is discussed to explain the role of feature selection in the classification and training phase of ML IDS. Finally, a discussion of the false and true positive alarm rates is presented to help researchers model reliable and efficient machine learning based intrusion detection systems.

## 1.0 INTRODUCTION

Computer networks have developed rapidly over the years contributing significantly to social and economic development. International trade, healthcare systems and military capabilities are examples of human activity that increasingly rely on networks. This has led to an increasing interest in the security of networks by industry and researchers. The importance of Intrusion Detection Systems (IDS) is critical as networks can become vulnerable to attacks from both internal and external intruders [1], [2].

An IDS is a detection system put in place to monitor computer networks. These have been in use since the 1980's [3]. By analysing patterns of captured data from a network, IDS help to detect threats [4]. These threats can be devastating, for example, Denial of service (DoS) denies or prevents legitimate users resource on a network by introducing unwanted traffic [5]. Malware is another example, where attackers use malicious software to disrupt systems [6].

Intrusion detection systems evolved as a response to these situations. Many IDS have been developed but have experienced the problem of false positive or false negative alarms. These are the false recognition of an attack by the IDS and increases the difficultly for network administrators to handle intrusion reports. Researchers in this field aim at developing IDS that have a high accuracy of detection and low false alarm rate [7]. Another problem of some existing IDS is their inability to detect unknown attack types. These IDS rely on the signatures of known attacks.

Human independent IDS that incorporate machine learning techniques have been developed as a solution to these problems. Machine learning IDS learns from normal traffic and abnormal traffic by training on a dataset to predict an attack by using classification. Several machine learning techniques have been successfully implemented as classifiers on IDS but present numerous flaws such as low throughput and high false detection rates [7].

The taxonomy (or classification) presented within this manuscript is highly relevant in the present study due to the highly diverse types of systems and attacks [8], [9]. The taxonomy will aid constructing two objectives: a clear description of the current state of IDS and the guide lines in which to explore the complexity of it [10], [11].

The paper aims to provide a clear description and guidelines needed to understand intrusion detection systems based on results from existing works in various papers. Its organisation gives a broader view of IDS and aims at addressing the concerns and solutions of shallow and deep learning IDS. The paper also compares previous works and their performance metrics.

The second objective of the paper is to present a survey and the classification of Intrusion Detection Systems, taxonomy of Machine Learning IDS and a survey on shallow and deep networks IDS.

## 2.0 INTRUSION DETECTION SYSTEMS

Intrusion detection systems are strategically placed on a network to detect threats and monitor packets. The IDS accomplishes this by collecting data from different systems and network sources and analysing the data for possible threats [12]. The functions of the IDS include offering information on threats, taking corrective steps when it detects threats and recording all important events within a network [13]. Figure 2 shows a model of Intrusion detection system

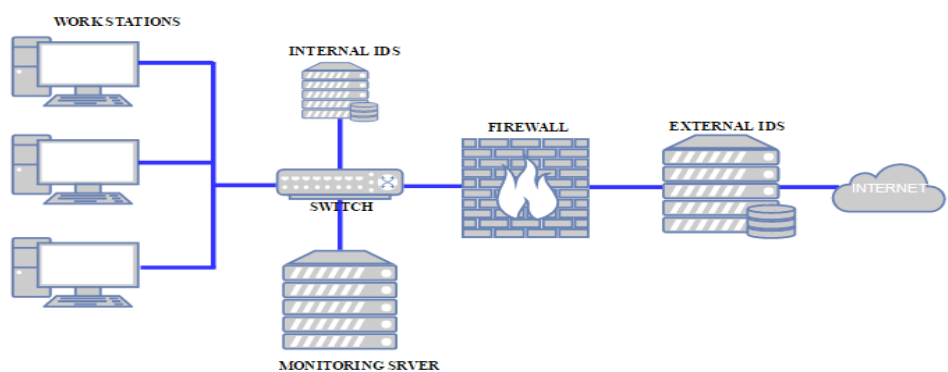

Figure 2 Intrusion detection system model

## 2.1 Classification of Intrusion detection system

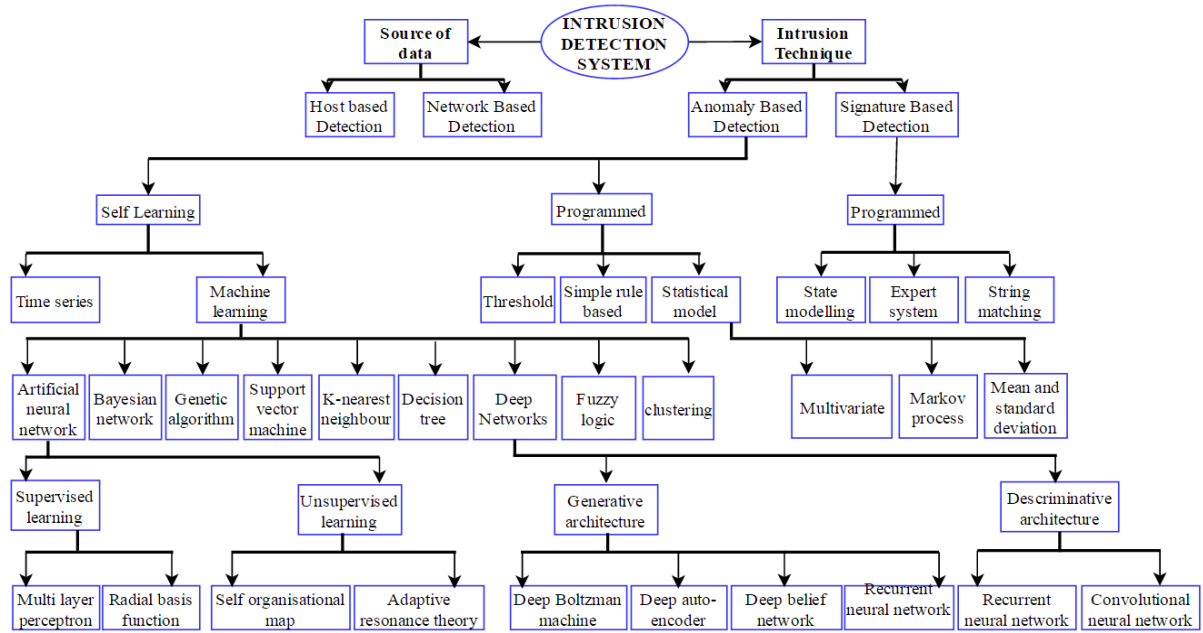

Figure 2.1 Classification of Intrusion Detection Systems

Different researchers have developed different classification representations. H.Debar *et al*[14], S.Axelsson *et al.[15]*, S.Amer *et al.*[16], C.Xenakis *et al.*[17], Hung-Ten Lia e*t al.*[18], have previously presented intrusion detection surveys and taxonomies. This research builds upon their work and introduces deep networks technique which all these referenced works do not describe. With the increasing value of big data, deep networks is an important element to capture in an IDS taxonomy  The taxonomy presented within this work provides a fine-grained overview of the different machine learning techniques for intrusion detection systems as shown in Figure 2.1. The detection mainly depends on the source of data and intrusion technique used.  The Source of data is the nodes that gather the information for analysis. The source can be either host or network based or both. These two sources are further discussed in section 2.1.1. The detection technique also differentiates between the anomaly based and signature based detection techniques. The different techniques that fall in the class of anomaly and signature based are important to develop a taxonomy that gives a broader view of techniques and the capabilities in detecting attacks. Section 2.1.2 and 2.1.3 explains anomaly based and signature based techniques in detail. These techniques operate either as self-learning or programmed. In Self-learning class, the system automatically learns whilst in programming class a user teaches the system.  The classification presented Fig.2.1 will help draw conclusions on detection accuracy and false alarm rates of IDS techniques.

### 2.1.1 Host Based IDS (HIDS) Vs. Network Based IDS (NIDS)
Host based IDS  were the first type of IDS to be implemented [19]. HIDS  are software based products installed on a host computer that analyse and monitor all traffic activities on the system application files and operation system [20], [21]. The traffic activities gathered by the system application files and system [20] are called the audit trails [22]–[24]. HIDS has an advantage of being able to detect threats from within by scanning through the traffic activities before sending and receiving data [25]. Its main disadvantage is that only monitors the host computer, meaning it has to be installed on each host [26].

Network based IDS are found at specific points on the network to capture and analyse the stream of packets going through network links, unlike the HIDS whose approach is to analyse each host separately [27] [28], [29]. It has the advantage of having a single system monitoring an entire network, saving time and cost of installing software on each host. NIDS main disadvantage is its vulnerability to any intrusion originating from the network targeting a system within the network.

Table I. Performance comparison of HIDS and NIDS [30]

| Performance in terms of: | Host-Based IDS | Network-Based IDS |
| --- | --- | --- |
| **Intruder deterrence** | Strong deterrence for inside intruders | Strong deterrence for outside intruders |
| **Threat response time** | Weak real time response but performs better for a long term attack | Strong response time against outside intruders |
| **Assessing damage** | Excellent in determining extent of damage | Very weak in determining extent of damage |
| **Intruder prevention** | Good at preventing inside intruders | Good at preventing outside intruders |
| **Threat anticipation** | Good at trending and detecting suspicious behavior patterns | Good at trending and detecting suspicious behavior patterns |

**2.1.2 Anomaly Based Detection**

Anomaly Based Detection is a behavioural based intrusion detection system. It observes changes in normal activity within a system by building a profile of the system which is being monitored [31], [32]. The profile is generated over a period of time when the system is established to have behaved normally [33]. One advantage is that it offers the ability to detect attacks which are new to the system [34]. Anomaly detection is categorised into two, based on the way a normal profile of a system is specified.

1) Self-learning – The self-learning system operate by example with a baseline set for normal operation. This is achieved by building a model for the underlying processes with the observed system traffic built up over a period of time [15]. Self-learning systems are sub-divided into the following main categories: time series model and machine learning. Section 3 discusses machine learning technique in detail.

    - **Time series model** takes into account the sequence of observation in order of succession occurring in intervals of uniformity. If the probability of occurrence of a new observation at a time is negligible then it is considered a change in normal behaviour. Time series has an advantage of observing trends of behaver over a period of time and flagging it, if it notices a change in normal behaviour. It is an effective model when attacks are sequential over a period of time [35]. This model has a disadvantage of being more costly

computationally [36]. Auto regressive moving average (ARMA) is an example of time series model used as IDS.

In Predictive Modelling for Intrusions in Communication Systems using generalised autoregressive moving average (GARMA) and ARMA models by [37] *et al.* fitted ARMA(1,1) and GARMA(1,2;$\delta$,1) time series models to 4 types of attacks (DoS, probe, U2R and L2R). The parameter estimation was done using Hannan-Rissanen algorithm, Whittle estimation and Maximum likelihood estimation and the point forecast obtained through Whittle estimation and maximum likelihood were close to the original value. The time series models were able to forecast the attack but the performance of GARMA (1, 2; $\delta$, 1) in attack detection was better.

.

2) Programmed − A programmed model is when a system needs either a user or an external person to teach the system to detect changes in behaviour. The user decides the extent of abnormal behaviour in the system and flags an intrusion threat [35]. The programmed models are grouped into four categories: threshold, simple rule based, and statistical models.

- **Threshold models** can be considered as the simplest programmed descriptive statistical detector [15]. By correlating and analysing statistical data, a user can program the system at a pre-defined threshold of alarm on a statistical variable. Careful selection of the threshold is required to minimise the false alarm rate. Instances where the threshold set too high leads to a risk of missing an alarm when an intruder is performing a malicious action [38]. A typical example is setting an alarm if three attempts to login to a system are unsuccessful [39].

  In anomalous payload-based network intrusion detection [40], Ke Wang *et al*. presented a payload base anomaly detection by demonstrating the effectiveness on 1999 DARPA dataset and a live dataset collected form the US Columbia CS department network. In training phase, a profile byte frequency distribution was computed and the standard deviation of application payload connected to a single host and port. During the detection phase the Mahalanobis distance was used to calculate the similarity of new data against the pre-computed profile. The detector compared this measure against a threshold and generated an alert when the distance of the new input exceeded this threshold. It recorded nearly 100% in accuracy with 0.1% false positive rate for port 80 traffic.

- **Simple rule based** is an approach for detection is to monitor events within a system that trigger a rule either than what is considered as a normal behaviour for the user. One limitation of this model is its failure to detect a threat that is not programmed as a rule in the system [41]. RIPPER (repeated incremental pruning to produce error reduction) is an example of rule based model that builds a set of rules to detect normal behaviour and abnormal behaviour. R.Naidu *et al*[42] used KDDCup '99 dataset to compare the performance of RIPPER rule, decision tree(C5) and support vector machine(SVM). The dataset was categorised into 3 (NORMAL, probe attacks and DoS attacks). The RIPPER rule algorithm went through two stages during the experiment. The first stage initialised the rule conditions and the second stage the rule optimisation. The algorithm obtained conditions in each rule to classify the testing data. The RIPPER rule recorded a total detection rate of 98.69%, C5 98.75% and SVM 98.63%.

- **Statistics model** collects data in a profile**.** Analysing the profile of normal statistical behaviour gives a description shown by the patterns from data to help make conclusions if an activity is normal or abnormal. The system then develops a distance vector for the observed traffic and the profile. An alarm is raised by the system when the distance is great enough [15],[14], [36], [43]. These models are sub categorised into four: mean/standard deviation, multivariate, Markov process and operational model. Dorothy Denning [44] discusses model based on the hypothesis that security violation can be detected by monitoring a system's audit records for changes in pattern. The model includes profiles for representing the behaviour of subjects with respect to objects in terms of metrics and statistical models and rules for acquiring knowledge about this behaviour from audit records and detecting anomalous behaviour.

    *a)* Mean, standard deviation and any other form of correlations are known as moments in statistics [35], [45]. A moment is said to be anomalous when events fall either above or below a set interval. Decisions are made taking system change into account by altering statistical rule set for the system [35]. Its advantage over the operational model is its ability to detect attacks without having prior knowledge of the normal activities to set its limits. Rather it learns form observations to determine its normal activities [35]. It is a complex model that has more flexibility than the threshold model. It does not require prior knowledge to determine the normal behaviour to help set pre-defined threshold. Varying the mean and standard deviations slightly changes the computation by putting extra weights on the more recent values [35]. A.Ashfag *et al*[46] proposed a standard deviation normalised entropy of accuracy hybrid method of intrusion detection. Two real traffic dataset were used. The endpoint dataset comprising 14months of traffic traces on a diverse set of 13 endpoints. The dataset was reduced to 6weeks of traffic for testing and training purposes. The endpoints were infected with different malware attacks. The second was attack data dataset was obtained from two international network locations at Lawrence Berkeley National laboratory, USA on three distinct days. Using 9 prominent classifiers on the two datasets showed 3%-10% increase in detection rate and 40% decrease in false alarm rate over the existing classifiers can be achieved with the proposed hybrid technique.

    **b)** Multivariate models are similar to the mean and standard deviation model [35], [47]. The multivariate models are based on correlations between two or more metrics. They use multiple variables to predict possible outcomes. For example the number of CPU cycles can be compared to how long a

login session is completed [47]. Theoretically this model could have a fine distinction over one variable [47]. In an approach by W.Sha *et al*.[48], multivariate time series and high-order Markov chain were taken into account along the detailed design of training and testing algorithms. The models were evaluated using DARPA dataset. Observing the multi order Markov chain showed that the relative positions between results from models of different orders provide a new effective indication for anomalies. To improve sensitivity, a combining multiple sequences as a multivariate one into a simple model was applied and proved that the return of values of the system calls also play an important role in detection.

c) Markov process has two approaches: Markov chains and hidden Markov models. The Markov model is a set of finite states interconnected going through a stochastic process to determine the topology and capabilities of a model [36]. Each stage in the process depends on the outcome of the previous stage. The anomalies are detected [35] by comparison of the associated probability recorded for the process with a fixed threshold. This gives it an advantage of detecting unusual multiple occurrence of events [47]. The hidden Markov model assumes the system to be a Markov process where stochastic processes with finite states of possible outcomes are hidden [36]. Ye Nong [49] presented an anomaly technique using Markov chain model to detect intrusion. In this model, Markov chain was used to represent a temporal profile of normal behaviour in a computer and network system. The Markov chain model of the normal profile is learned from the historic data of the system's normal behaviour. The observed behaviour of the system is analysed to see if the Markov chain model of the normal behaviour supports the observed behaviour. A low probability of support indicates an anomalous behaviour meaning an intrusion. The technique was implemented on the Sun Solaris system and distinguished normal activities from attacks perfectly.

**2.1.3 Signature Based Detection**

Signature based detection defines a set of rules used to match the patterns in the network traffic. If a mismatch is detected it raises an alarm [50]. It has an advantage of being able to detect attacks giving a low false positive detection ratio [51] It has a drawback of being able to detect only attacks known to the database [19]. Signature based detection systems are programmed with distinct decision rules. The rules set for detection are coded in a straight forward manner to detect intrusion. They are programmed in four categories: state modelling, expert system, and string matching.

- State modelling is the encoding of attacks as a number of different states in a finite automaton. Each of these attacks has to be observed in the traffic profile to be considered as an intrusion. They occur in sub-classes as time series models [15]: the first is state transition which was proposed by Porras *et al*. which uses a state transition diagram to represent intrusion [52]. The approach in [52]models intrusion as a series of state transitions which are described as signature action and states descriptions. State diagrams were written to correspond to the states of an actual computer system. The basis of a rule based expert system to detect intrusions was formed by these diagrams. The intrusions are in the form of a simple chain transitioning from start to end. The second is the petri-net where states form a petri-net. They form a tree structure where the transition states are not in any order [15][53].

- Expert systems contain a set of rules used to describe attacks scenarios known to the system. The given rules that describe the attack scenarios are often forward-chaining systems. A production based expert system tool has been used since they best handle systems with new events entering into the system. The size of the rule based increases as the execution speed increases since the rule will go through a longer list. It is however vulnerable to attacks not known to the set rules [2]. P.Zhisong *et al.*[54], presented an intrusion detection system model based on neural network and expert system. The aim of the experiment was to take advantage of classification abilities of neural network for probe and Dos attacks and expert based for U2R and R2L attacks. KDD Cup'99 dataset was employed in the experiment. Expert system was able to improve the detection accuracy using the detection rules: ALERT UDP ENET any <-> HNET 31337 MESSAGE:"BO access" DATA: |ce63 d1d2 16e7 13cf 3ca5 a586|".

- String matching is a process of knowledge acquisition just as Expert system but has a different approach in exploiting the knowledge [14]. It deals with matching the patterns in the audit event generated by the attack but not involved in the decision making process [47]. This technique has been used effectively commercially as an IDS [55], [56]. It is noted that not all signature based IDS can be represented by a simple pattern which gives it a limitation [57]. T.Sheu *et al.*[58] proposed an efficient string matching algorithm with compact memory as well as high worst-case performance. A magic number heuristic based on the Chinese remainder theorem was adopted. The algorithm significantly reduced the memory requirements without bringing complex processes. The latency of off-chip memory references was drastically reduced. It was concluded the algorithm gives a cost effective and an efficient IDS.

## 2.1.4 Discussion

A summary table of intrusion techniques, source of data, applications and data used in selected reviewed papers between 2014 and 2016 is shown in Table II.

Table II. Summary table of selected reviewed papers.

| Authors/Year | Year Published | Intrusion Technique | Source of Data | Application | Data Set |
| --- | --- | --- | --- | --- | --- |
| **Chetna and Harsh** [59] | 2014 | Anomaly | NIDS | Cloud computing | Real Life |
| **M. Ali Alheeti** *et al.* [60] | 2016 | Anomaly | NIDS | Transport | Real Life |
| **J. Hong** *et al.* [61] | 2014 | Anomaly | NIDS | Substation | Real Life |
| **R. Mohan** *et al.* [62] | 2015 | Signature | HIDS/NIDS | Cloud computing | Real Life |
| **S. Vasudeo** *et al.* [63] | 2015 | Anomaly/ Signature | HIDS/NIDS | Computer systems | Real Life |
| **W.Haider** *et al.* [64] | 2015 | Anomaly | HIDS | Cloud computing | ADFA-LD and KDD98 |
| **H. Toumi** *et al.* [65] | 2015 | Signature | NIDS | Cloud computing | Real Life |
| **M. Guerroumi** *et al.* [66] | 2015 | Signature | HIDS | Internet of things (IoT) | Real Life |
| **N. Dipika** *et al.* [67] | 2015 | Anomaly | NIDS | Information systems | KDD99 Cup |
| **W.Haider** *et al.* [68] | 2015 | Anomaly | HIDS | Cyber space | KDD98 and UNM |
| **N.Aissa** *et al.* [69] | 2015 | Anomaly | NIDS | Computer systems | KDD99 |
| **H. Omessaad** [70] | 2015 | Signature | HIDS | Cloud computing | Real Life |
| **X. Lin** *et al.* [71] | 2015 | Signature | NIDS | Transport | Real Life |
| **S. Banerjee** *et al.* [72] | 2015 | Signature/ Anomaly | NIDS | Computer systems | Real Life |
| **P. Satam** [73] | 2015 | Anomaly | NIDS | Telecommunication | Real Life |

Recent works have shown that works are still on-going using the Intrusion technique and source of data techniques. S.Vasudeo *et al.* [63] presented a hybrid of Signature and Anomaly with a hybrid of data source on a real life dataset. S.Banerjee *et al.*[72]also combined signature and anomaly technique on real life dataset. Other referenced works the table applied single techniques on real life datasets and KDD Cup datasets. The application of these techniques cut across all areas to determine their effectiveness in detecting intrusion.

## 3.0 MACHINE LEARNING TECHNIQUES

Anomaly detection systems are human-independent. They detect anomalies by revealing abnormal characteristics in a system over a period of time [44]. The effectiveness of this technique is its capabilities to differentiate between normal and abnormalities within a network.

Machine Learning (ML) can provide IDS methods to detect current, new and subtle attacks without extensive human-based training or intervention. It is defined as a set of methods that can automatically detect patterns to predict future data trends [74],[75]. Whilst a large number of machine learning techniques exist, the fundamental operation of all of them relies upon optimal feature selection. These features of are the metrics which will be used to detect patterns and trends. For example, one feature of a network is the packet size: machine learning techniques may monitor the packet size over time and generate distributions from which conclusions may be drawn regarding an intrusion. This section reviews the feature selection of machine learning IDS and classification of machine learning techniques used as IDS in Figure 3.

### 3.1 IDS feature selection

Machine learning classification involves two phases: the classification phase and training phase. The training phase learns the distribution of the features and during the classification phase the learned features is applied as a normal profile where any abnormality will be detected [76], [77]. A.K. Jain *et al.* [78] developed a model of statistical pattern recognition as shown in Fig.3. The test and training data are normalised in the processing unit as well as removing noise form the data. In the training phase, the feature extraction unit extracts a representation feature set from the processed training data which are used in training a classifier. In the classification phase, the trained classifier is applied to assign the test data to the selected features from the training phase [77].

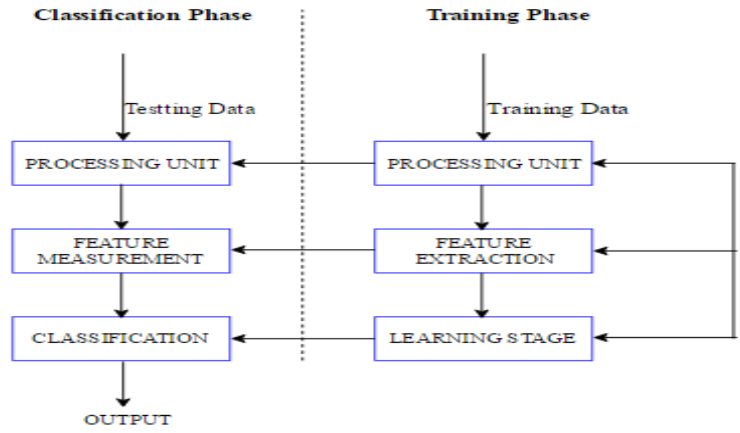

Figure 3 Model of Machine learning classification process

A high quality of training data is required to achieve the best performance of ML IDS. The training data thus contain both normal and abnormal patterns [76]. Features are the important information extracted form from raw data and are important in classification and detection which influence the effectiveness of ML IDS. In [79], C. Kruegel *et al*. created attacks containing system call sequences similar to normal system call. Kruegel *et a*l. addressed the issue by analysing on the system call arguments instead of finding the relations between sequences of actions [76]. Table III. Shows the features extracted by Kruegel *et al*. for four different arguments.

Table III. Extracted features by Kruegel *et al*. [76] for four different arguments

| No. | Feature Description |
|---|---|
| 1 | String length |
| 2 | String character |
| 3 | Structural inference |
| 4 | Token finder |

In [80], V. Mahoney *et al.* extracted features by analysing each header-field of network packets and flows. The proposed approach det6ected anomalies in a network based on 33 fields of IP, UDP, TCP, ICMP and Ethernet protocols. Table IV shows some of the extracted features.

Table IV. Extracted features based on analysis of each header field

| No. | Feature Description | No. | Feature Description |
|---|---|---|---|
| 1 | IP Source | 10 | IP protocol |
| 2 | IP Destination | 11 | IP Length |
| 3 | TCP Source Port | 12 | IP Type of Service |
| 4 | TCP Destination Port | 13 | IP Header Length |
| 5 | UDP Source Port | 14 | IP Time to Live |
| 6 | UDP Destination Port | 15 | Ether Source |
| 7 | UDP Length | 16 | Ethernet Size |

| 8 | ICMP Code | 17 | Ethernet Destination |
|---|-----------|----|--------------------|
| 9 | ICMP Type | 18 | Ethernet Protocol |

In [81], W. Lee *et al.* extracted features from TCP/IP connections. Experiment was conducted on 1998 DARPA dataset where a set of basic features from domain knowledge was extracted. Most of the basic features could only be extracted after the TCP connection was terminated leading to a delay in the detection phase [76]. The 1998 DARPA dataset is categorised into five. These are NORMAL and four different types of attack. The attacks are: Dos attacks, probe attack, User to root (U2R) attack and Remote to local (R2L) attacks. Dos attack which denies or prevents legitimate users resource on a network or system by introducing useless or unwanted traffic. Probe is an attack which scans through a computer network to make a profile of information for future attacks. U2R is attacks use a local account from a remote machine to gain access to the targets system due to vulnerabilities in its operating system. R2L attacks are initiated to gain unauthorised access to root privileges from outside. Table V. illustrates the basic features of individual TCP connections by W. Lee *et al.*

Table V. Basic features extracted from individual TCP connections

| No. | Feature Name | Feature Description | Type |
|-----|--------------|---------------------|------|
| 1 | Duration | Length of connection in seconds | continuous |
| 2 | protocol_type | Type of the protocol used, e.g.. up,tcp, etc. | discrete |
| 3 | Service | Network service on the destination, e.g.http, telnet, etc. | discrete |
| 4 | scr_bytes | Number of data bytes from source to destination | continuous |
| 5 | dst_bytes | Number of data bytes form destination to source | continuous |
| 6 | Flag | Normal or error status of network | discrete |
| 7 | Land | 1 if connection is from/to same host/port; 0 otherwise | discrete |
| 8 | wrong_fragment | Number of wrong fragments | continuous |
| 9 | Urgent | Number of wrong fragments | continuous |

Table VI. Traffic features extracted from a 2second time window

| No. | Feature Name | Feature Description | Type |
|-----|--------------|---------------------|------|
| 10 | Count | Number of connections to the same host as the current connection in the past 2seconds | continuous |
| 11 | serror_rate | Percentage of connections having "SYN" error | continuous |
| 12 | rerror_rate | Percentage of connections having "REJ" error | continuous |
| 13 | same_srv_rate | Percentage of connections having same service | continuous |
| 14 | diff_srv_rate | Percentage of connections having different service | continuous |

| 15 | srv_count | Number of connections to the same service as the current 1connection in the past 2seconds | continuous |
|---|---|---|---|
| 16 | srv_serror_rate | Percentage of connections having "SYN" errors | continuous |
| 17 | srv_rerror_rate | Percentage of connections having "REJ" errors | continuous |
| 18 | srv_diff_host_rate | Percentage of connections to different host | |

Table VII. Traffic features extracted from a window of 100 connections

| No | Feature Name | Feature Description | Type |
|---|---|---|---|
| 19 | dst_host_count | Count for destination host | continuous |
| 20 | dst_host_serror_rate | Percentage of connections having "SYN" errors | continuous |
| 21 | dst_host_rerror_rate | Percentage of connections having "REJ" errors | continuous |
| 22 | dst_host_same_srv_rate | Percentage of connections having same service | continuous |
| 23 | dst_host_diff_srv_rate | Percentage of connections having different service | continuous |
| 24 | dst_host_srv_count | Number of connections to the same destination port | continuous |
| 25 | dst_host_srv_serror_rate | Percentage of connections having "SYN" errors | continuous |
| 26 | dst_host_srv_rerror_rate | Percentage of connections having "REJ" errors | continuous |
| 27 | dst_host_srv_diff_host _rate | Percentage of connections to different host | continuous |
| 28 | dst_host_same_src_port _rate | Percentage of connections to the same source port | continuous |

Table VIII. Traffic features extracted within a connection suggested by domain knowledge

| No. | Feature Name | Feature Description | Type |
|---|---|---|---|
| 29 | Hot | Number of "hot" indicators | continuous |
| 30 | num_failed_logins | Number of failed login attempts | continuous |
| 31 | logged_in | 1 if successfully login; 0 otherwise | discrete |
| 32 | num_compromised | Number of "compromised" conditions | continuous |
| 33 | root_shell | 1 if root shell is obtained; 0 otherwise | discrete |
| 34 | su_attempted | 1 if "su root" command attempted; 0 otherwise | discrete |
| 35 | num_root | Number of "root" accesses | continuous |

| 36 | num_file_creations | Number of file creation operations | continuous |
|---|---|---|---|
| 37 | num_shells | Number of shell prompts | continuous |
| 38 | num_access_files | Number of operations in access control files | continuous |
| 39 | num_outbound_cmd | Number of outbound commands in an ftp session | continuous |
| 40 | is_hot_login | 1 if the login belongs to the "hot" list; 0 otherwise | discrete |
| 41 | is_guest_login | 1 if the login is a "guest" login; 0 otherwise | discrete |

W. Lee *et al.* discovered many attacks such as R2L and U2R were in the payloads of packets. This prompted a proposal to combine features extracted from the payload with domain knowledge. They called the features "time based traffic" features for connection records. Features 10 to 18 in table VI are extracted using a 2second time window with 11 to 15 having same host connection and 16 to 18 having same service connection, 19 to 28 in Table VII are extracted from a window of 100 connections with 20 to 24 same host connection and 25 to 28 same service connection. 29 to 41 in Table VIII are features extracted from connections suggested by domain knowledge [77]. Both tables provide a detailed insight mapping the most current and meaningful features for machine learning intrusion detection systems.

## 3.2 Classification of Machine Learning Techniques

This section gives an extensive classification of ML techniques. Each of the techniques is described in detail and how these techniques have been applied as IDS. Figure 3.1 shows the classification of Machine learning IDS.

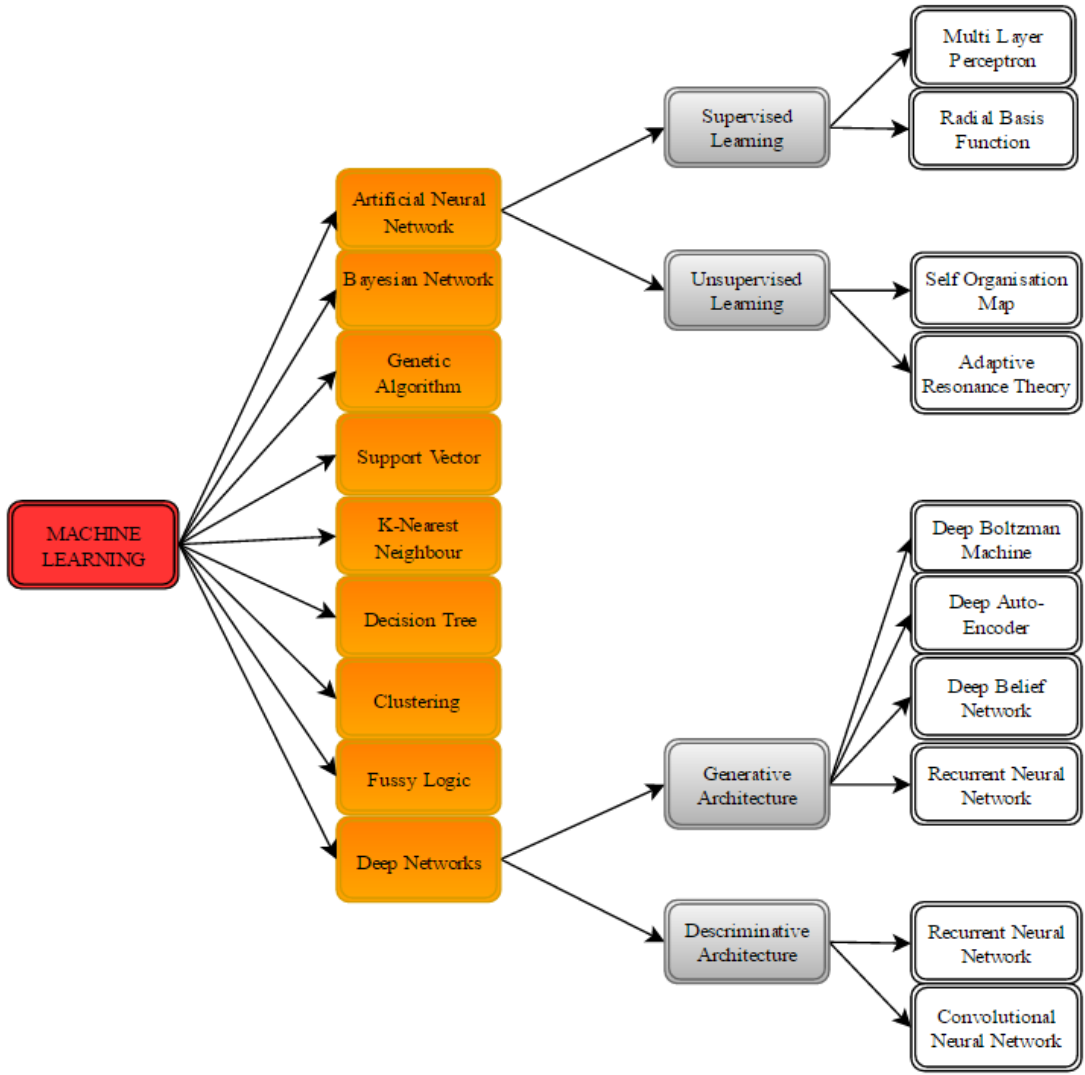

Figure 3.1 ML based techniques

- **Bayesian Networks** are graphical modelling tools used to model the probability of variables of interest. They are directed acrylic graphs where each node represents a discrete random variable of interest. Each node contains the states of the random variable in tabular form representing the conditional probability table (CPT) which specifies conditional probabilities of the domain variable with other connected variables. The CPT of each node contains probabilities of the node being in a specific state given the domain variable states. The existence of the relationship between the nodes of the domain variable and connected variables in a Bayesian networks show the direction of causality between other connected variables. Thus the connected variables are causally dependent on the ones represented by the domain variable. Given a set of discrete random variable $X = \{x_1, x_2 \ldots x_n\}$, the joint probability of the variable can be computed based on Bayes Rule as:

$P(x_1, x_1 \ldots x_n) = \prod_{t=1}^{n} P(x_i \backslash p_a(x_i))$ (1)

where $p_a(x_i)$ represents the specific values of the variables in the domain variable node of $x_t$ [82], [83]. This technique has generally been used as an intrusion detection system.

A. Onik *et al.*[84] Conducted an experiment using Bayesian networks on NSL-KDD dataset containing 25,192 records with 41 features. Apart from the normal class label it contained 4 more class of attacks known as Dos, U2R, R2L and probe attacks. The filter approach of feature selection was used to reduce the dataset features from 41 to 16 important features. Bayesian model was built and proved to predict attacks with superior overall performance accuracy rate of 97.27% keeping the false positive rate at a lower rate of 0.008. The model as compared to Naïve Bayes, K-means clustering, decision stamp and RBF network recorded 84.86%, 80.75%, 83.31% and 91.03% respectively in terms of accuracy.

An experiment by M. Bode *et al.*[85] Analysed the network traffic in a cyber situation with Bayesian network classifier on the KDD Cup'99 dataset with 490,021 records. The data set was made up of 4 types of attacks (DoS, U2R, R2L and probe). In this experiment, they adopted the risk matrix to analyse the risk zone of the attacks. The risk analysis adopted showed DoS was most frequent attack in occurrences. The results showed Bayesian network classifier is a suitable model resulting in same performance level classifying the DoS attacks as association rule mining. Bayesian network classifier outperformed Genetic Algorithm in classifying probe and U2R attacks and classified Dos equally.

- **Genetic Algorithm (GA)** is an adaptive search method in a class of evolutional computation using techniques inspired from convolutional biological process. The principle is based on a stochastic global [25] search method initialising with a random generation of chromosomes. The chromosomes are called population. They evolve through selection, crossover and mutation as shown in Figure 3.2. Each chromosome represents a problem to be solved and encoded as strings. The positions of the chromosomes are commonly represented as binary (0, 1) or as a list of integers. These positions sometimes referred to as genes keep changing at each initialisation. The solution created during each generation is based on an evaluation function. The selection is thus based on the chromosome fitness level [86], [87].

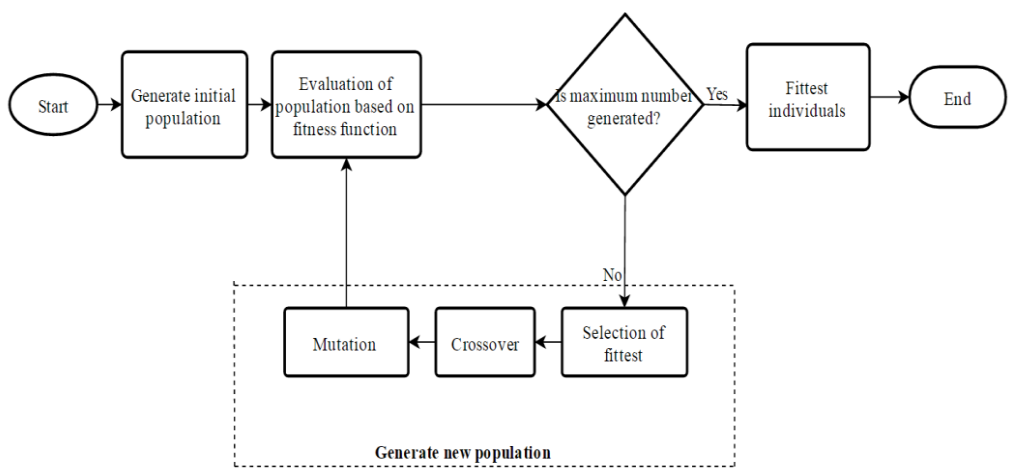

Figure 3.2 Flow chart of Genetic Algorithm

Tao Xia *et al.* [88] developed a hybrid method based information theory and GA. Information theory was used to filter out the most important features out of 41 features in KDD'99 dataset with 494021 records. Linear rule was used initially to classify normal and abnormal before GA to obtain the appropriate classification. In the detection of Dos, U2R, R2L and probe attacks, the information theory and GA hybrid method recorded 99.33%, 63.64%, 5.86% and 93.95% as detection rates respectively. The detection rates were compared to Ctree and C5. Ctree recorded 98.91%, 88.13%, 7.41% and 50.35% whilst C5 recorded 97.1%, 13.2%, 8.4% and 83.3% respectively for the attacks.

In layered approach for intrusion detection systems based GA, M. Padmadas *et al.*[89] Proposed a method to overcome the weakness in a single layer intrusion detection system. The layered approach is based on GA with the four layers corresponding to four groups of attacks (probe, DoS, U2R and R2L). Each layer is trained separately with a number of features where the layer acts as a filter to block any malicious activity. In layered approach there is no mathematical approach to calculate the filter parameters for the attacks. This paper presented GA approach in calculating the filter parameters making the system more secure. The model efficiently detected R2L attack and recorded an accuracy of 90% in detection.

- **Support Vector Machines (SVM)** is a machine learning algorithm that learns to classify data using points labelled training examples falling into one or two classes. The SVM algorithm builds a model that can predict if a new example falls into one category or the other [90],[25], [91] . Figure 3.3 shows a hyperplane defined by $(w, b)$, where $w$ is a weight and $b$ bias constructed in a finite space of the training sample $N$ with points:

$$\{(x_1, y_1)(x_2, y_2 \ldots (x_N, y_N))\} \qquad (2)$$

Where $x_i \in R^d$ and $y_i \in \{1, -1\}$. This is conducted since in general the larger the margin, the lower the generalisation error of the classifier [92].

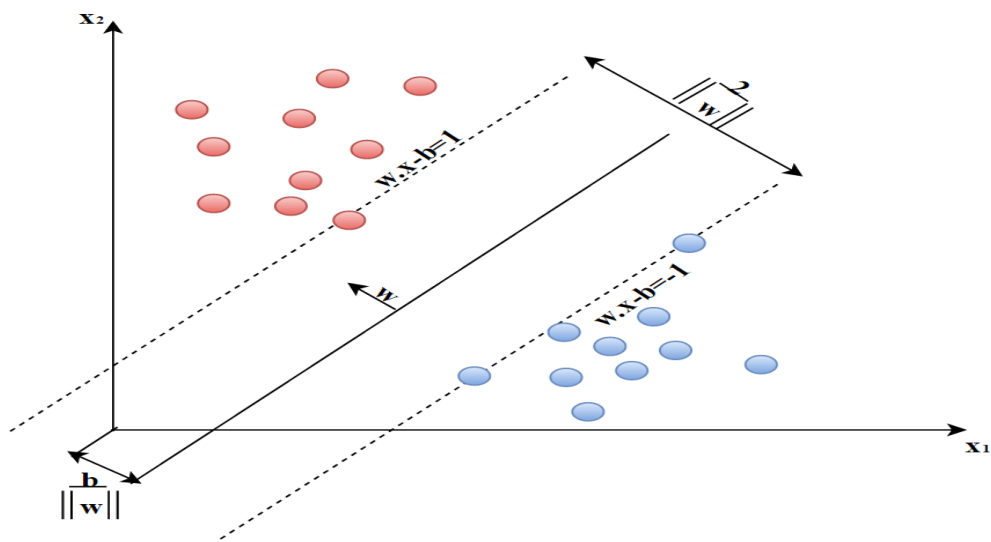

Figure 3.3 Maximum-margin hyper plane and margins for an SVM trained with samples from two classes.

B. Senthilnayaki *et al.*[93] Used GA pre-processed KDD Cup 99 dataset in a pre-processing module for data reduction since it was complex to process the dataset with all 41 features. GA was used to select 10 features out of 41 features present in the KDD Cup 99 dataset and applied SVM for classification. The experiment was carried out with 100,000 records from the dataset out of which 95% was used as training data and remaining 10% as test data. The classification process continued till a 10 fold cross validation was done for results verification. The SVM classified four different attacks (DoS, probe, U2R, R2L attacks). The performance of SVM classifier was compared with four other classifiers as shown in table IX.

Table IX. SVM performance analysis SVM, GMM, Naive, MLP and linear algorithm

| Classified Algorithm | Accuracy% (41 feature) | Accuracy% (10 features) |
|---|---|---|
| **Gmm** | 30.5 | 45.07 |
| **Naive** | 45.17 | 55.75 |
| **MLP** | 55.24 | 67.87 |
| **Linear** | 65.47 | 76.75 |
| **SVM** | 82.45 | 95.26 |

Another experiment by L.Teng *et al*. [94] proposed a novel method integrating PCA and SVM by optimising the kernel parameters using automatic parameter selection technique. The experiment was performed on KDD Cup'99 dataset containing five categories of traffic (normal, DoS attack, R2L attack, U2R attack and probe attack). Each network record had 41 features of which 7 were discrete and 34 continuous features. C parameter for RBF kernel of SVM was optimised by the proposed automatic parameter reduction along with cross validation to reduce the training and testing time to give better accuracy in detecting attacks.

- **K-Nearest Neighbour (K-NN)** is a fundamental technique for sample classification. This technique is known to be non-parametric and highly efficient in classification [95]. It evaluates the class labels of the test samples [96] based on the majority of test sample neighbours. The parameter $k$ is determined by the user. Based on the test sample, $k$ numbers of training points are determined by taking the closest distance to the test sample. The prediction of the test sample is the $k$ nearest neighbours [97].

A hybrid method of intrusion detection system by Y.Canbay *et al.*[96] proposed combination of K-NN and GA algorithm. They tested the hybrid algorithm on KDD Cup 99 dataset labelled out of five classes; normal, probe attack, Dos attack, R2L attack and U2R attack. The dataset was reduced to 1000,2000,3000,4000 and 5000 records respectively with 19 features. GA was used to select the $k$ nearest neighbour for K-NN classifier. Three experiments were performed to conclude on the hybrid method. Each experiment used 10 fold cross validation with different $k$ values. The values in in terms of accuracy were compared with the conventional K-NN. The proposed hybrid method proved better than conventional K-NN in all values of $k$ used in the experiment.

On the same dataset Q.Zeng *et al*. [98]compared the performance of K-NN and SVM model and RIPPER method in detecting attacks. The multi attribute decision was adopted in this experiment. In the classification of an unknown document vector $X$, $k$–nearest neighbour algorithm ranks the document's neighbour among the training document vectors and uses the class labels of the $k$ most similar neighbours to predict the class of the new document. The similarity in each neighbour to $X$ is used to determine the classes of the neighbour where the similarity is measured by the Euclidean distance between two document vectors. With this adoption they categorised each new program behaviour in the dataset into either normal or attack class. Each system call was treated as a word and each process as a document. $k$ was varied between 15 and 35 till an optimal value of 19 found. The classification was performed with K-NN and SVM model. The Hit rate was compared to the RIPPER method. The results showed 97.26% accuracy rate and 6.03% false alarm rate for K-NN and SVM model whilst the RIPPER method gave 87.26% accuracy rate and 8.6% false positive rate.

- **Decision Tree (DT)** algorithm learns and models a data set in classification problems. It classifies new data set according to what it has learnt from previous data set [99]. It uses a well-defined criterion in the selection of best features of each node tree during their construction. A decision tree model has a root node linking to different nodes as attribute data deciding the path for each node [100]. Decisions are made by comparison of previous data and marked as leaves [101]. A common decision tree approach is the C4.5 algorithm.

  In network intrusion detection system using decision tree (J48) by S.Sahu *et al*. [102] a labelled data set called Kyoto 2006+ was used. The data consist of 24 features, 14 of which was extracted form KDD Cup 99 dataset and an additional ten important features. The Perl language was used to extract 15 features from the dataset for the experiment. The sample dataset contained 134665 records; 44257 normal, 86649 known attacks and 3759 unknown attacks. Decision tree (J48) built using WEKA 3.6.10 tool was used to classify normal, known attacks and unknown attacks in the network packets. The results showed the decision tree generated classified 97.23% correctly and 2.67% incorrectly. The simulation results showed decision tree can classify unknown attacks as well.

  T.Komviriyavut *et al.*[101]presented two intrusion detection techniques which were decision tree (C4.5) and RIPPER rules to test an online dataset (RLD09 dataset). RLD09 dataset was collected from actual environment and refined to have 13 features. The dataset is categorised into three; normal, DoS attack and Probe attack. The experimental data set had 3,000 records of three types of unknown probe attacks each with 1,000 attack records. Initial experiment with known attacks on dataset showed a total detection rate of about 98% for both decision tree and RIPPER rule.

  A second experiment on the same data set with three unknown probe attacks (advance port scan, Xmas tree and ACK scan) was performed. The decision tree maintained its detection rate of 98% whilst the RIPPER rule degraded to about 50% in average.

- **Fuzzy logic (FL)** concept is derived [103] from the fuzzy sets theory which deals with approximately reasoning with uncertainty and imprecision [104] by human being. The features this technique which handles real life uncertainty makes it attractive for anomaly detection. Intrusion detection involves the classification of a normal class and an abnormal class. The well-defined nature of the two classes makes this computation paradigm a helpful one.

A.Toosi *et al.*[105]introduced a new approach using different soft computing techniques into classification system to classify abnormal behaviour form normal depending on the attack type. This work investigated neuro-fuzzy networks, fuzzy inference and GA on KDD Cup 99 10% dataset. The dataset had a distribution of normal and four different attacks (probe, DoS, R2L and U2R). A set of parallel neuro-fuzzy classifiers were used initially for classification. The fuzzy inference system was based on the output of the neuro-fuzzy classifier determining normal or abnormal activity. The best results were attained by optimising the structure of the fuzzy decision tree with GA.

- **Clustering** is the division of data with no apparent distinguished differences into groups by functions of two or more parameters. Each group is called a cluster and have no similarity. The greater the dissimilarity the better the grouping. Their class labels are always unknown. K-means is an example of clustering algorithm used as IDS [106],[107]. K-means algorithm makes use of the Euclidean distance to calculate the data and the cluster centre. The aim is to have a minimum distance within a cluster and achieve a maximum distance between clusters by the minimisation of an objective function [108].

Li Jun Tao *et al.*[109]proposed a k-means clustering with dynamic adjustable number of cluster. They introduced an algorithm that uses an improved Euclidean distance formula to calculate the distance between the data and cluster centre by automatically adjusting the number of cluster when the distance is more than the threshold. This experiment was performed on KDD Cup'99 dataset containing four attacks (DoS, Probe, R2L and U2R). The total number of dataset records was 72471. The improved k-means algorithm recorded 77 clusters with a detection rate of 90% and false positive rate of 15%. The results compared to the traditional k-means were 108 clusters, detection rate of 85% and false positive rate of 42%.

## 3.3 Analytical Comparison of ML Techniques

The various ML techniques discussed in the above section are different in their capabilities to classify attacks. Table X. analyses the advantages and disadvantages in the performance of ML techniques.

Table X. Advantages and disadvantages of ML techniques

| ML Technique | Advantages | Disadvantages |
|---|---|---|
| **Bayesian Network** | The graphical representation gives it an advantage of breaking complex problems into different smaller models. | Slow in classifying data sets with many features. |
| **Genetic Algorithm** | -Uses a technique which is inspired by convolutional biological process.<br>-It has the capability of solving optimisation problems during classification. | Gets stack in local optima (overfitting) |
| **Support** | -The algorithm is simple to analyse | -Selection of kernel function is not straight |

| | | | |
|---|---|---|---|
| **Vector Machine** | mathematically.<br>-All computations are performed in space using kernels giving it an edge to be used practically. | forward.<br>-Slow in training and requires more memory space. | |
| **K-Nearest Neighbour** | Easy to implement and can solve multi-class problems | -Slow in training and requires large memory space.<br>-It is computationally complex because to classify a test sample involve the consideration of all training samples. | |
| **Decision tree** | -It has a unique structure therefore easy to interpret.<br>-It has no limitation in handling high dimensional data sets. | -If trees are not pruned back it causes overfitting.<br>-Type of data must be considered when constructing tree.<br>(i.e. Categorical or numerical) | |
| **Fuzzy Logic** | -It is based on human reasoning concepts which are not precise.<br>-It gives a representation of uncertainty. | -It's construction has a high level of generality there by high consumption of resource. | |
| **K-means Algorithm** | Simple to implement and effective | -The outcome of clustering depends on how cluster centres are initialised to specify k value.<br>-The algorithm works for only numerical data. | |

## 3.4 ML Hybrid Techniques

ML Intrusion detection systems have generally been used to detect attacks but in recent times, using two or more different techniques to form a hybrid has improved the overall performance. Table XI. Shows accuracy of detection, type of attacks and data set for selected reviewed ML IDS papers from 2011-2016.

Table XI. ML IDS accuracy of detection and Data set used

| Authors(s)/Year | ML Technique | Attack Type(s) | Data Set | Accuracy |
|---|---|---|---|---|
| S.Chordia *et al.* [107] 2015 | K-means +KNN+DT | R2L,U2R,DoS and probe | KDDCup99 | 96.55% |
| P. Jongsuebsuk *et al.* [110] 2013 | FL+GA | Dos and Probe | Real Life | 97% |
| B.Senthilnayaki *et al.* [93] 2015 | GA+SVM | Dos<br>Probe<br>U2R<br>R2L | KDDCup99 | 99.15%<br>99.08%<br>97.03%<br>96.50% |
| B. Masduki *et al.* [7] 2015 | SVM | R2L | KDDCup99 | 96.08% |
| A.Enache *et al.* [111] 2015 | SVM+BAT Algorithm | Malicious | NL-KDD | 99.38% |
| S. Akbar *et al.* [112] 2012 | GA | DoS,R2L,U2R and Probe | KDDCup99 | 92.6% |
| A. Aziz *et al.* [113] 2013 | DT | Dos<br>Probe | NSL-KDD | 82%<br>64% |
| Chao Lin *et al.* [114] 2015 | Cluster centre+ K-NN | Probe<br>Dos | KDDCup99 (6 dimensional data set) | 99.98%<br>99.99% |
| A.Aburomman *et* | SVM+K-NN | DoS,R2L,U2R and | KDDCup99 | 87.44-91.69% |

| al. [115] 2016 | | Probe | | |
|---|---|---|---|---|
| Shi-Jin Horng et al. [116] 2011 | SVM + hierarchical clustering algorithm | DoS,R2L,U2R and Probe | KDDCup99 | 95.72% |
| E. Hodo et al.[117] 2016 | ANN | DDoS/Dos | Real Life | 99.4% |

As demonstrated in Table XI, the majority of Intrusion Detection Systems using machine learning have tested their work against the KDDCup99 dataset or the NSL-KDD, in contrast only two recent studies tested their machine learning systems against real network data.

### 3.5 Binary Classification metrics

The effectiveness of prediction by the ML algorithm which is either 1 or 0 is based on confusion matrix prediction outcome as shown Table IX. The outcomes are True Negative (TN), True Positive (TP), False Positive (FP) and False Negative (FN) [118], [119].

Table IX. Confusion Matrix

| | | Predicted Class | |
|---|---|---|---|
| | | Negative Class(Normal) | Positive Class(Attack) |
| **Actual Class** | Negative Class(Normal) | True Negative(TN) | False Positive(FP) |
| | Positive Class(Attack) | False Negative(FN) | True Positive(TP) |

- True Negative (TN): a measure of the number of normal events rightly classified normal.
- True Positive (TP): a measure of attacks classified rightly as attack.
- False Positive (FP): a measure of normal events misclassified as attacks.
- False Negative (FN): a measure of attacks misclassified as normal.

The following are the basic metrics used to calculate the performance of ML IDS:

True negative rate (Specifity) $= \frac{TP}{FP+TN}$ (3)

True positive rate (Sensitivity) $= \frac{TP}{TP+FN}$ (4)

False positive rate (Fallout) $= \frac{TP}{TP+FN} = 1 - Specifity$ (5)

False negative rate (Miss Rate) $= \frac{FN}{FN+TP} = 1 - Sensitivity$ (6)

Precision $= \frac{TP}{TP+FP}$ (7)

$$\text{Recall} = \frac{TP}{TP+FN} \tag{8}$$

$$\text{Overall accuracy} = \frac{TP+TN}{TN+TP+FN+FP} \tag{9}$$

A commonly used ML IDS metric is detection rate. This is defined as the number of data examples correctly classified divided by the test examples.

## 4.0 ARTIFICIAL NEURAL NETWORK AND DEEP NETWORKS IDS

This section reviews artificial neural network (ANN) and Deep learning which use the computational intelligence approach to detect attacks.

### 4.1 Artificial Neural Network

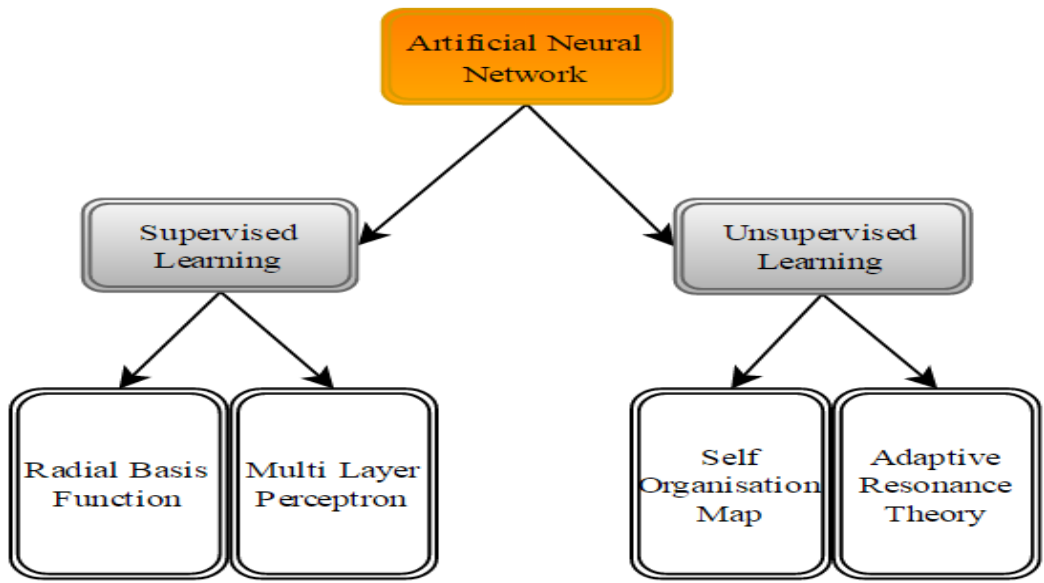

Figure 4 Classification of Artificial Neural Network

Artificial neural network consist of information processing elements known to mimic neurons of the brain. ANN is categorised into supervised and unsupervised learning. Figure 4 shows the types of ANN.

### 4.1.1 Supervised Learning
In supervised learning, the neural network is provided with a labelled training set which learns a mapping from inputs $x$ to outputs $y$ given a labelled set of inputs-output pairs

$$d = \{(x_i, y_i)\}_{i=1}^{N} \tag{10}$$

Where $d$ is called the training set and $N$ is the number of training examples. It is assumed that $y_i$ is a categorical variable from some infinite set $y_i \in \{1 \ldots C\}$ [120]. Two types of supervised learning algorithms are used to train a neural network for intrusion detection.

- Multilayer Perceptron (MLP) is a feedforward neural network. The structure of MLP consists of one or more nodes as the inner layer between the input and output nodes as shown in Figure 4.1. The most common technique used to train the MLP neural network is the Back Propagation hence the name MLP-BP. The construction of the MLP-BP neural network is by putting layers of non-linear elements to form complex hypotheses. The more stages that are added (nodes) the more advance the hypotheses. Each node takes an element of a feature vector. The output nodes give an output of two classes (normal and attack). The interconnection between the nodes is associated with scalar weights with an initial weight assigned to the connection. During training, the weights are adjusted. Evaluating the hypotheses is done by setting the input modes in a feed-back process and the values are propagated through the network to the output. At this stage the gradient descents is used so as to push the error in the output node back through the network by a back propagation process in order to estimate the error in the hidden nodes. The gradient of the cost – function can thus be calculated [99], [121].

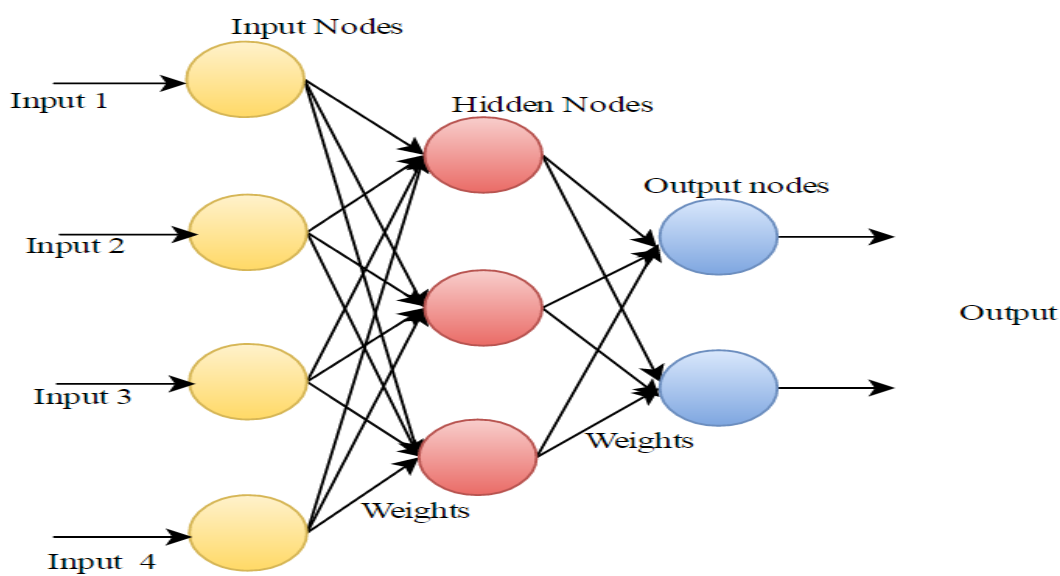

Figure 4.1 Three layer neural network

In [122], P. Barapatre *et al.*[122]experimented with neural network having an input layer, one hidden layer and one output layer. In this experiment, the input node had 41 features from KDD Cup'99 dataset. The output node was to classify normal or attack present in dataset. If the output node is '0' then it is labelled "normal" and if '1' it is labelled "attack". The learning rate was initially set to 0.1 and the system retrained to reduce learning rate. A sigmoid activation function was used and values corresponding to the weights randomly initialised between -1 and -. The system was trained for each category of attacks (DoS, Probe, U2R and R2L attacks) to determine the performance of the algorithm on individual attacks. Table IIX shows the experimental results. It was concluded MLP-BP neural network detected DoS and Probe attacks more accurately than U2R attacks. It observed an increase in rate of classification as rate of learning decreases giving rise to a slow convergence. Also reducing the rate of learning and the sum squared error (SSE) values and re-training the network showed an improvement in rate of detection.

Table IIX Detection rate comparison for different attack types using MLP

| Attack type | Detection rate % | False positive rate % | False negative rate % |
|---|---|---|---|
| **DoS** | 99.75 | 4.78 | 0.24 |
| **Probe** | 98.16 | 1.33 | 1.83 |
| **U2R** | 87.09 | 31.03 | 12.9 |
| **DoS-Probe** | 99.33 | 18.5 | 0.66 |
| **R2L** | 98.99 | 9 | 1.01 |
| **Overall** | 81.96 | 8.51 | 18.03 |

- Radial Basis function (RBF) is another feed forward neural network. It classifies by taking a measurement of the distance between the inputs and the centre of hidden neurons [118]. Figure 4.2 is an RBF architecture showing the input nodes, one hidden nodes and output. Each RBF has different parameters with an input vector. The network output is thus a linear combination of the radial basis function's output. The input and hidden nodes weights are always 1 since the transfer function of the network is a Radial basic function. This allows an adjustment on the weight between the hidden nodes and the output [123].

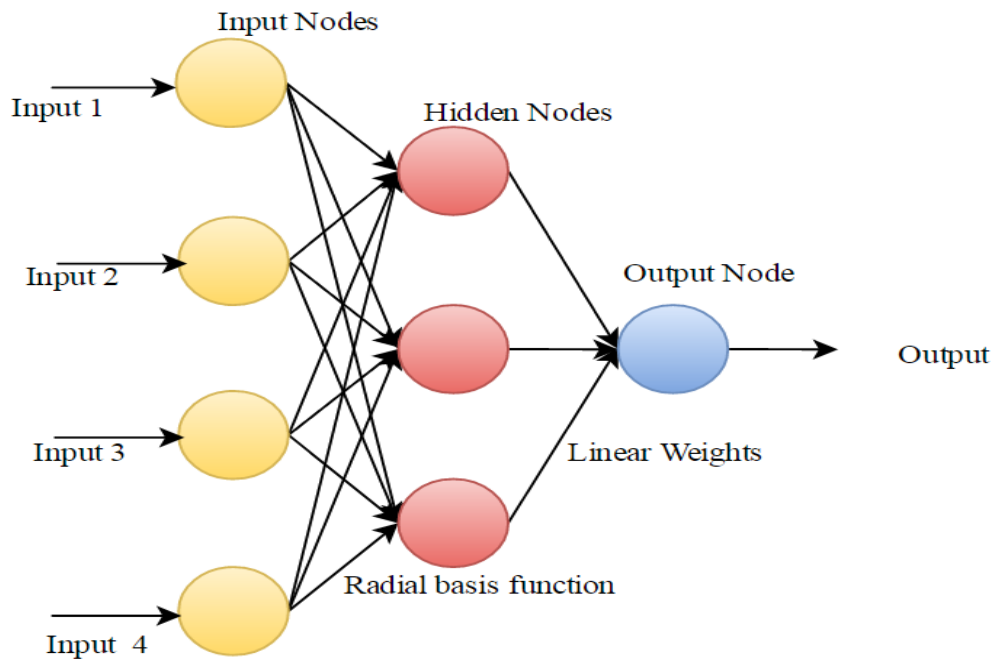

Figure 4.2 Radial basis function architecture

C.Zhang *et al*.[124] compared the performance of RBF and MLP-BP in detection detecting four different types of attacks on KDD Cup'99 dataset. Both training and testing data contained 1000 records which were selected from the dataset. In this experiment each sample was unique with 34 numerical features and 7 symbolic features. The symbolic features were converted to ASCII numbers before being used as training or testing data. The results showed RBF achieved better performance than MLP-BP with a detection rate of 99.2% and false positive rate of 1.2%. MLP-BP showed a detection rate of 93.7% and false positive rate of 7.2%.

Ju Jiang *et al*. [125] applied RBF and Back propagation algorithm (BPL) to both misuse and anomaly detection. The experiment was performed on KDD Cup '99 dataset containing 4,900,000 records categorised into 5 (normal, probe attacks, DoS attacks, R2L attacks and U2R attacks). The misuse detection network had 41 input features, 4 hidden nodes and 4 output nodes representing normal probe, DoS and R2L classes respectively. The anomaly network also had 41 input features, 1 hidden node and 1 output node to classify normal and attacks behaviour. The experimental results showed in misuse detection, RBF based IDS performed similarly to BPL based IDS. However RBF used a shorter time in training as compared to BPL based IDS and needed to adjust its decision thresholds. In anomaly detection, the BPL based IDS had to adjust itself output threshold manually according to the characteristics of the training dataset to achieve best performance. In anomaly detection RBF based IDS out performed BPL based IDs.

Bi Jing *et al*. [123] compared RBF and MLP-BP using a processed KDD Cup '99 dataset by converting al strings to numeric, reducing the dimension of the dataset and determining the rational value domain. The features were reduced from 41 to 31 to train both RBF and MLP-BP. The RBF structure used had one hidden layer and output neuron being the weighted sum of all the output items of the hidden layer. The simulation results showed RBF network is better than MLP-BP in its property of having a more regular out with shorter training time and better accuracy in attack detection.

### 4.1.2 Unsupervised learning

In unsupervised learning, the neural network is only provided with input data without conceptualising the output: it discovers patterns within the data autonomously. The data yet to be discovered is called unlabelled data [120]. Two typical unsupervised learning are Self-Organisation Maps (SOMs) and Adaptive Resonance theory (ART).

- Self-Organisation Maps

    SOMs transform the input of a network into two dimensional feature maps based on the topological properties of SOMs. The computation of feature maps is by Kohonen unsupervised learning. The two dimensional feature maps are neurons represented by coloured squares showing the weights corresponding to each neuron. The inputs are grouped based on their similarity. The more features mapped, the bigger the coloured square. The quality is determined by the back ground colouring of the clusters. Anomaly events can thus be identified by analysing the normal and abnormal events from the mapping [118], [126].

    SOMs are widely used anomaly detection systems. P. Lichodajewski *et al*. in [127] applied SOMs as a Host based intrusion detection system. In this work SOM was trained with an explicit coding of data and gave a clear clustering of abnormal behaviours.

    In [128], H. Gunes Kayacik *et al.* investigated and demonstrated with hierarchical SOM architecture with two basic feature sets, one limited to 6 basic features with the others containing all 41-features. The results gave a false positive rate of 1.38% and detection rate of 90.4%.

V. Kumar *et al* in [129] presented a unified framework on SOM. Their approach detected attacks on a mobile ad-hoc network (MANET) using different parameters. Their experimental results were found to be better than other neural network approaches in terms of detection rate and false alarm rate.

- Adaptive resonance Theory (ART)

  ART as a basic theory is an unsupervised learning model but as a hybrid it performs supervised learning. It functions as a pattern recognition and predictive tool. As an unsupervised learning model, ART-1, ART-2, ART-3 and fuzzy Art makes the list. The supervised models are made up of ARTMAP, Fuzzy ARTMAP and Gaussian ARTMAP [118]. The model compares an input vector to a single neuron's weight (weight vector) [118], [126].

  In [130], M. Chauhan *et al*. presented a novel technique by uploading the weights of the network based on sign of an evaluation function [131]. An appropriate evaluation function was selected by utilizing the probabilities of various states. It concludes that if the weights in a modified ART network are oscillating, it implies an intrusion and thus the ART network has the capability of detecting any intrusion.

  A hybrid method by P. Somwang *et al.* in [132] used principal component analysis (PCA) and fuzzy adaptive resonance theory to identify different attacks on KDD Cup '99 data set. The results showed a high performance of detection rate of 96.13% and a false alarm rate of 3.86% of anomaly intrusion detection.

### 4.1.3 Summary

I. Ahmad *et al*. in [133] evaluated SOMs, ART and MLP-BP in terms of main criteria and sub-criteria using analytic Hierarchy. Evaluation based on main criteria investigated less overhead, maturity, competency, performance and suitability. Sub-criteria investigated cost effective, time saving, detection rate, minimum false positive, minimum false negative, handling varied intrusion and handling coordinated intrusion. Different radar graphs with distinct colours were used for the analysis.

Analysing MLP-BP and SOM showed MLP-BP had a better detection rate, minimum false positive, minimum false negative, time saving and cost effective. In the case of less overhead and capability of handling coordinated and varied intrusion, MLP-BP was not as good as SOM. Analysing MLP-BP and ART showed the same results.

A comparison of ART and SOM showed better results for SOM in terms of less overhead and handling coordinated intrusion. In another scenario ART was better than SOM in terms of detection rate, minimum false negative, maturity, time saving and cost effective.

In conclusion, a hybrid ANN approach was found to be the most suitable intrusion detection system in terms of detection rate, false positive, false negative, cost and time saving.

### 4.2 Deep Networks

This section focuses on networks that look like the multilayer perceptron (MLP) but have a different architecture. The difference between the MLP and the deep networks is their training procedures. This is a class of ML techniques where classification is conducted by training data with many layers in hierarchical networks with unsupervised learning. Deep networks are inspired by the architecture depth of the brain. In 2006, Hinton et al. from the university of Toronto came up with Deep Belief Network (DBN) [134]. They trained data with an algorithm that greedily trains layer by layer using unsupervised learning for each layer of Restricted Boltzmann Machine (RBM) [135]. After this discovery by Hinton et al. other deep networks have been introduced using the same principle have been successful in classification task [136]. Deep networks IDS can be classified based on how the architectures and techniques are being used. Figure 4.3 shows a classification of Deep networks IDS.

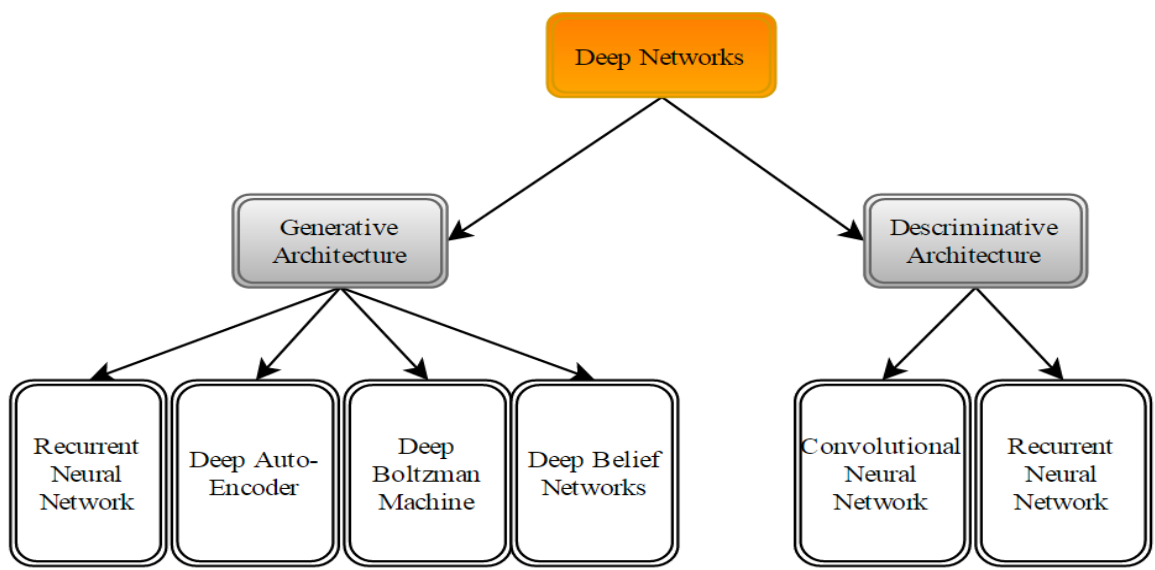

Figure 4.3 Classification of deep learning IDS

### 4.2.1 Generative Architecture

Generative models are also referred to as graphical model because they depict independence/dependence for distribution. They are visualised as graphs that have nodes representing random variables and arcs show the relationship between random variables that can have millions of parameters to graphically represent the given system [136], [137]. The joint statistical distribution of the variables can be written as products of the nodes and their associated variables [136], [138]. The graphical models have hidden variables that cannot be observed. Generative models are associated with supervised learning since their training does not depend on the labels of the data. For classification purposes the models goes through a pre-training stage (unsupervised learning). During this process, each of the lower layers are trained separately from the other layers which allows the other layers to be trained in a greedily layer by layer from bottom to up. All other layers are trained after pre-training [136]. The sub-classes of generative models are Recurrent Neural Network (RNN), Deep Auto-Encoder, Deep Boltzmann Machine (DBM) and Deep Believe networks (DBN).

- Recurrent Neural Network (RNN)

    RNN is a class of deep networks that are either considered supervised or unsupervised learning with an input sequential data whose length could be as large as its depth [139]. The RNN model architecture is a feedback loop linking layer by layer with the ability to store data of previous input increasing the reliability of the model [140]. There are two types on the RNN in terms of architecture: Elman and Jordan RNNs. The Elman model has a simple feedback looping layer by layer. The Jordan model has

a feedback looping all neurons within a layer to the next layer. There exists also a feedback connecting a neuron to itself. The ability of the Jordan RNN to store information in the neurons allows it to train less input vector for classification of normal and abnormal patterns with high accuracy [140]. Figures 4.4 and 4.5 are simple architectures of the Elman RNN and Jordan RNN showing the context unit known to store information of the previous output of hidden layer.

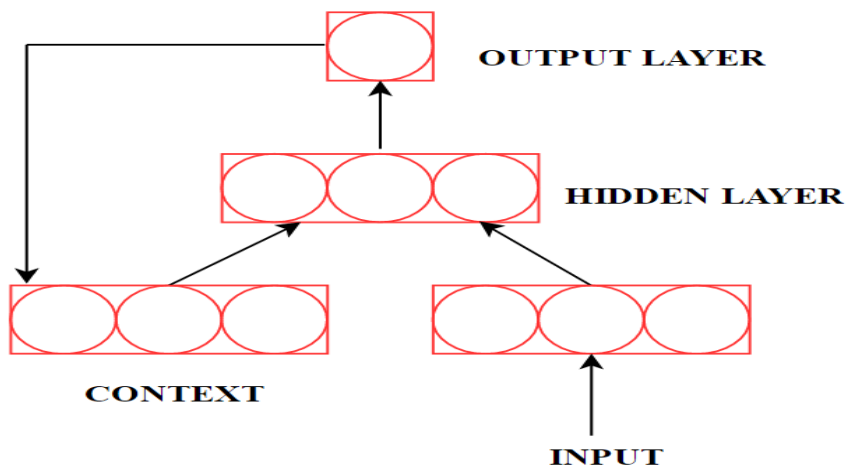

Figure 4.4 Jordan Network

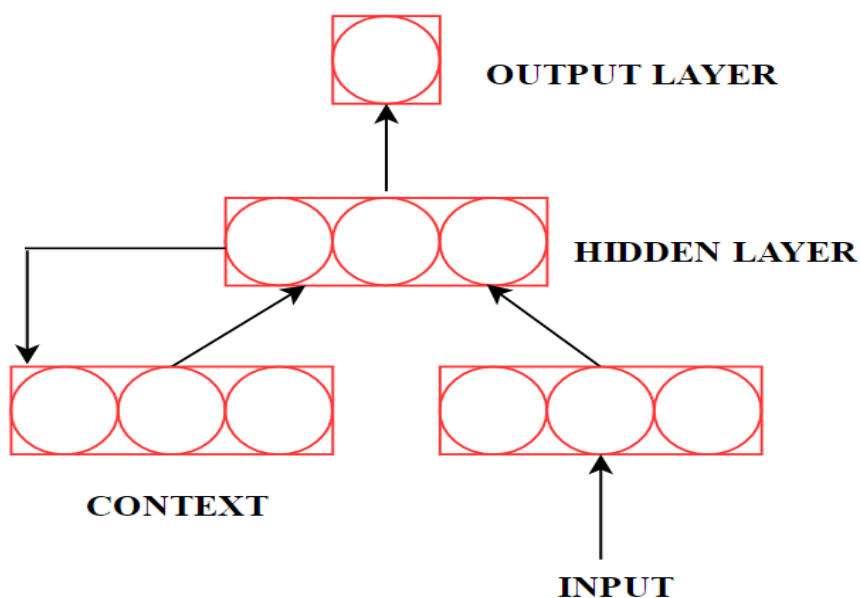

Figure 4.5 Elman Network

A study by K. Jihyun *et al*. [141] applied long short term memory (LSTM) architecture to RNN and trained the IDS using KDD Cup '99 dataset. By comparing the accuracy with other IDS classifiers, LSTM-RNN recorded 96.93% accuracy with a detection rate of 98.88%. Although the FAR was slightly higher than the others, they concluded its overall performance was the best.

- Deep Auto-Encoder
These are energy based deep models classified as generative models in their original form. It comes in different forms mostly also generative models. Other forms are stacked auto encoder and de noising auto encoder a [138].

An Auto-Encoder becomes deep when it has multiple hidden layers. It is made up of an input layer unit representing the sample data, one or two hidden layer units where the features are transformed and then mapped to the output layer unit for reconstruction. Training the auto-encoder gives it a "bottleneck" structure where the hidden layer becomes narrower than the input layer to prevent the model from learning its identity function [120].

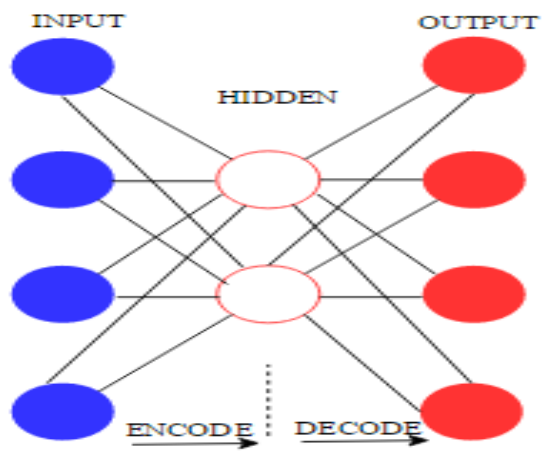

Figure 4.6 Deep auto encoder

Unfortunately, the deep auto encoder when trained with back propagation has not been a success the evaluation gets stuck in local minima with a minimal gradient signal when trained with back propagation. Pre-training a deep auto encoder using the greedy layer wise approach by training each of the layers in turns has proven to alleviate the backpropagation problems [120], [139],[142].

In [143], B. Abolhasanzadeh proposed an approach to detect attacks in big data using deep auto encoder. The experiment was conducted on NSL-KDD data set to test the method of applying bottle neck features in dimensionality reduction as part of intrusion detection. The results in terms of accuracy rate out performed PCA, factor analysis and Kernel/PCA. It was concluded; the results recorded in terms of accuracy makes this approach promising one for real world intrusion detection.

- Deep Boltzmann Machine(DBM)

  DBM is a unidirectional graphical model. Currently there exist no connection between units on the same layer but between the input units and the hidden units. DBM when trained with a large supply of unlabelled data and fine-tuned with labelled data acts as a good classifier [136]. Its structure is an offspring of a general Boltzmann machine (BM) which is a network of units based on stochastic decisions to determine their on and off states [138]. BM algorithm is simple to train but turns to be slow in the process. A reduction in the number of hidden layers of a DBM to one forms a Restricted Boltzmann Machine (RBM) [139]. DBM when trained with a large supply of unlabelled data and fine-tuned with labelled data acts a good classifier. Training a stack of RBM with many hidden layers using the feature activation on one RBM as the input for the next layer leads to the formation of Deep Believe Network (DBN).

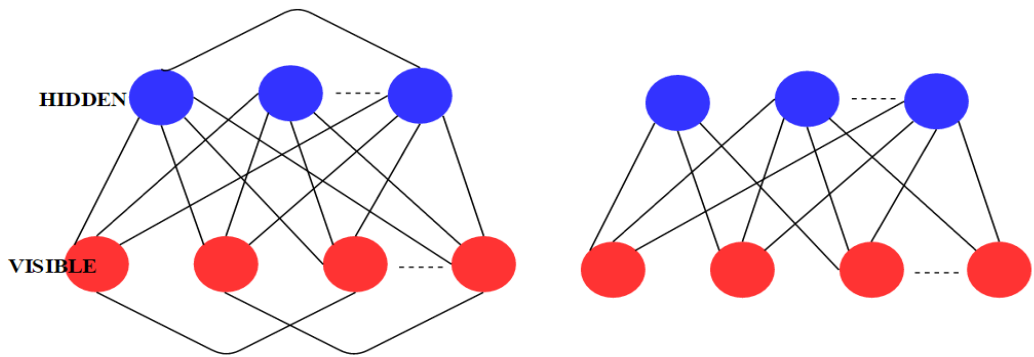

Figure 4.7 Left: A Boltzmann machine with stochastic on and off hidden features and a variable layer representing a vector of stochastic on and off states. Right: An RBM with no visible to visible units connected and no hidden to hidden units connected.

U. Fiore *et al*. in [144]explored RBM in anomaly detection by training a network with real world data traces from a 24hour work station traffic. This experiment was to test the accuracy of RBM to classify normal data and data infected by bot. a second experiment trained RBM with KDD Cup '99 data set and tested against real world data. To randomize the order of test data, the experiment was repeated 10times. The experiment confirmed testing a classifier in two different networks training data affects the performance. They suggested the nature of anomalous traffic and normal traffic should be investigated.

- Deep Belief Networks (DBN)

  DBN uses both unsupervised pre training and supervised fine-tuning techniques to construct the models. Figure 4.8 shows a DBN which is made up of a stack of Restricted Boltzmann Machines (RBMs) and one or more additional layers for discrimination task. RBMs are probabilistic generative models that learn a joint probability distribution of observed (training) data without using data labels. Once the structure of a DBN is determined the goal for training is to learn the weights *w* between layers. Each node is independent of other nodes in the same layer given all nodes which gives it the characteristic allowing us to train the generative weights of each RBM [120]. It then goes through a greedy layer by layer learning algorithm which learns each stack of RBM's layer at a time. In Figure 4.8 left, the top layers in red form a RBM and the lower layers in blue form directed sigmoid believe network [145].

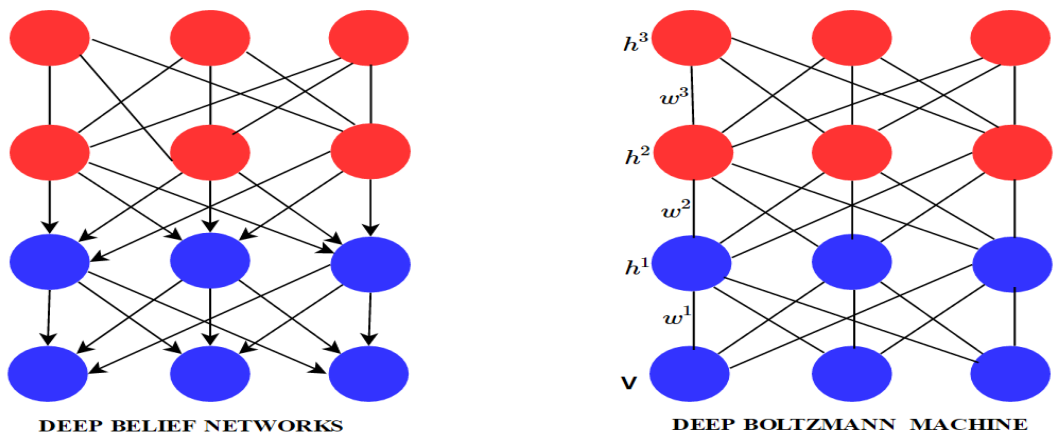

Figure 4.8 Three layer Deep Belief network and three layers deep Boltzmann machine

DBN has been trained as a classifier by N. Gao *et al*. [146] to detect intrusion by comparing the performance to SVM and ANN. The classifiers were trained on KDD data set. The authors proved that deep learning of DBN can successfully be used as an effective ID. They concluded the greedy layer by layer learning algorithm when used to pre-train and fine-tune a DBN gives a high accuracy in classification. The results showed that DBN recorded the best accuracy of 93.49%, a TP value of 92.33 and FP of 0.76%.

Z. Alom *et al*.[147] also exploited the DBNs capabilities to detect intrusion through series of experiments. The authors trained DBN with NSL-KDD data to identify unknown attack on it. They concluded by proposing DBN as a good IDS based on an accuracy of 97.5% achieved in the experiment. This results was compared with existing DBN-SVM and SVM classifiers which it out performed.

**4.2.1 Discriminative Architecture**

The discriminative architecture uses discriminative power for classification by characterising the posterior distributions of classes conditioned on the input data. Recurrent neural network and convolutional neural network are two types of discriminative architecture.

- Recurrent neural network (RNN) uses discriminative power for classification when the model's output is an explicit labelled data in sequence with the input data sequence. To train RNN as a discriminative model, training data needs to be pre-segmented and a post-processing to transform the output to a labelled data [138].
- Convolutional neural network

  A convolutional neural network (CNN) is a type of discrimination deep architecture with one or more convolutional and pooling layers in an array to form a multilayer neural network [139],[148],[149]. In general, convolutional layers share many weights followed by sampling of the convolutional layer's output by the pooling layer which results in some form of translational invariant properties [139].

  CNN has fewer parameters as compared to other connected networks with the same number of hidden units which gives it an advantage of easier training [148]. CNN architecture is that of multi-layer perceptron [150], and are variant of MLP which are inspired biologically. Hubel and Wiesel worked on the cat's visual cortex and deduced that visual cortex is made up of an arrangement of cells in a complex manner. These cells are sensitive to small sub-regions of the visual field known as the receptive field. These fields are positioned to shield the entire visual field to enable the cell behave just like a local filter over the input space [151]. The series of layers making up a CNN architecture are the convolutional layer, max pooling layer and the fully connected layer [150], [152]. The convolutional layer is made up of neurons forming a rect6angular grid where previous layers are made of neuros shaped as a rectangular grid. The rectangular grid neurons are connected to each other with the inputs from previous rectangular units through a set of weights known as filter banks [149], [152]. These weights for the rectangular units do not change for every rectangular grid of neuron to form convolutional layers. In architectures where the convolutional layer is made up of different grids [152], each grid uses a different filter bank [149], [153]. Each convolutional layer is followed by a pooling layer which merges subsets of the convolutional layer's rectangular block by taking sub-samples to give an output of the block. The pooling can be done in several ways such as [152] computing the maximum or average or a learned linear summing of neurons in the blocks. Some blocks turn to be shifted more than a row or a column and in turn feed in an s input to neighbouring pooling units. This causes a reduction in the dimension of the system design and thus causing variations to the input [149], [153]. The final stage which has several convolutional and max-pooling layers non-linearly stacked in the neural network form a fully connected layer of network [152]. The connectivity which allows the set of weights of the filter banks to be trained easily [149], [152].

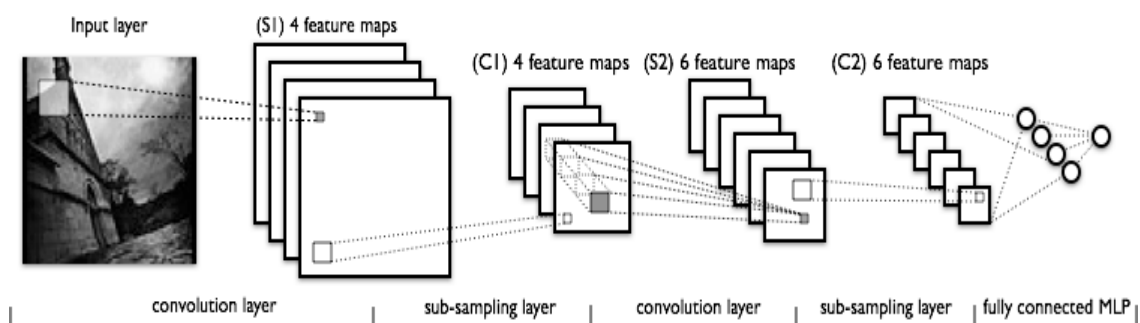

Figure 4.9 Image of CNN showing layers making up architecture [154]

## 5.0 Conclusion

Shallow and deep networks intrusion detection systems have gained a considerable interest commercially and amongst the research community. With advancement in data sizes, intrusion detection systems should have the characteristics to handle noisy data with high accuracy in detection with high computational speed. This paper gives an overview of the general classification of intrusion detection systems and taxonomy with recent and past works. This taxonomy gives a clear description of intrusion detection system and its complexity.

Current studies of deep learning intrusion detection systems have been reviewed in this paper to help address the challenges in this new technique still in its early stages in intrusion detection. In particular recent papers have been reviewed in this work considering all the machine learning techniques including the single and hybrid techniques.

The scope of the work on classifying intrusion detection systems, reviewing the various methods of detecting anomaly, performance of these methods were based on past and recent works revealing the advantages and disadvantages of each of them.

The focus of the paper on shallow and deep networks described experiments comparing the performance of these learning algorithms. The experiments demonstrated deep networks significantly outperformed the shallow network in detection of attacks.

To the best of our knowledge CNN has not been exploited in the field of intrusion detection but has proven to be a good classifier. DBN is also new in its exploitation in this field and experimental works are still in progress to determine the reliability of these learning algorithms to detect attacks.

Signature based technique have been in use commercially but have not been able to detect all types of attacks especially if the IDS signature list did not contain the right signature.

Research work is in progress experimenting new approaches to test the reliability and efficiency of knowledge based and behavioural approaches in intrusion detection.